\documentclass[aps]{revtex4}
\usepackage{amssymb}
\usepackage{amsmath}

\input{tcilatex}

\begin{document}

\title{Potential for regulatory genetic networks of gene expression near a
stable point }
\author{Ming-Chang Huang$^{1}$, Yu-tin Huang$^{2}$, Jinn-Wen Wu$^{3}$, and
Tien-Shen Chung$^{1}$ }
\affiliation{${\ }^{1}$Center for Nonlinear and Complex Systems and Department of
Physics, Chung-Yuan Christian University, Chung-Li, 32023 Taiwan\\
${\ }^{2}$C.N. Yang Institute for Theoretical Physics, State University of
New York, Stony Brook, 11790-3840 USA\\
${\ }^{3}$Department of Applied Mathematics, Chung-Yuan Christian
University, Chung-Li, 32023 Taiwan }

\begin{abstract}
A description for regulatory genetic network based on generalized potential
energy is constructed. The potential energy is derived from the steady state
solution of linearized Fokker-Plank equation, and the result is shown to be
equivalent to the system of coupled oscillators. The correspondence between
the quantities from the mechanical picture and the steady-state fluctuations
is established. Explicit calculation is given for auto-regulatory networks
in which, the force constant associated with the degree of protein is very
weak. Negative feedback not only suppresses the fluctuations but also
increases the steepness of the potential. The results for the fluctuations
agree completely with those obtained from linear noise Fokker-Planck
equation.
\end{abstract}

\date{June 25, 2007}
\maketitle
\preprint{YITP-SB-07-33}


A regulatory network of gene expressions consists of a group of genes which
co-regulate one another's expressions. Such networks provide a fundamental
description of cellular function at the DNA level. Recently, the advance of
experimental techniques in constructing synthetic networks with the ability
of monitoring them has provided some essential elements, such as switch\cite%
{ptashne,hasty,gardner} and oscillator\cite{elowitz,atkinson}, for the
design of biological circuits. In modeling the dynamics of a regulatory
network, rate-equation approach is often used; the approach reflects the
macroscopic observation with deterministic nature. However for systems with
small molecular number, intrinsic fluctuations become important. The
noise-induced effect may be incorporated into the framework by employing the
master equation and then proceeding via stochastic Monte Carlo simulations.
In general, master equation is discrete in nature. By using the technique of 
$\Omega $-expansion\cite{kampen}, we may convert master equation to
continuous Fokker-Planck equation which, then, is managed analytically by
various approximations. Significant progress has been made along this line
in understanding the regulation mechanism. One of the noticeable examples is
the auto-regulatory networks of a single gene for which, the protein,
encoded in the gene, serves as the regulator of itself through either
negative or positive feedback. Such autoregulation is a ubiquitous motif in
biochemical pathways. It was demonstrated by Becskei and Serrano that an
autoregulatory network with negative feedback may gain stability\cite%
{becskei}. Further analyses was given by Thattai and van Oudenaarden\cite%
{thattai} and by Ozbudak et al.\cite{ozbudak}, and the results indicate that
noise is essentially determined at the translational level and negative
feedback can suppress the intrinsic noise. Moreover, Tao and Tao et al. used
the linear noise Fokker-Planck equation to study the fluctuations and
obtained the results consistent qualitatively with previous works\cite%
{tao,tao-c}. \ 

One may conclude from the results above that the intrinsic noise associated
with a genetic network is closely related to its regulation scheme. This
Letter then attempts to provide a physical picture on this relation via the
establishment of a mechanical analogous system. To achieve this, we first
construct the solution of non-equilibrium steady state for the Fokker-Planck
equation near a stable point. Then, the potential of the system, defined as
the negative of the logarithm of the solution, can be first approximated as
an harmonic oscillator potential . Subsequently, we introduce a measure for
the steepness of the potential near a stable point and give the exact
relations between the force constants of coupled oscillators and the
correlations of fluctuations. Thus, the physical property of a regulation
scheme can be revealed from the corresponding mechanical analogue specified
by the force constants of coupled oscillators. This paper starts with the
general construction for a $d$-dimensional regulation network, followed by
the explicit calculations of auto-regulatory networks.

Consider a $d$-dimensional regulatory network of gene expression. The
network is specified by the macroscopic rate equations

\begin{equation}
\overset{\cdot }{x}_{i}=f_{i}\left( x\right)  \label{eq01}
\end{equation}%
with $i=1$, $2$, $...$, $d$, and the drift force $f_{i}$ defined as 
\begin{equation}
f_{i}\left( x\right) =R_{i}\left( x\right) -\phi _{i}x_{i}.  \label{eq02}
\end{equation}%
Here, $x^{\tau }=\left( x_{1}\text{, }x_{2}\text{, }...\text{, }x_{d}\right) 
$ with the superscript $\tau $ for the transpose of a vector, $x_{i}$
represents the concentration of mRNA\textit{\ }or protein, the function $%
R_{i}\left( x\right) $ describes the synthesis or feedback regulation of
molecule $i$, and the constant $\phi _{i}$ denotes the degradation rates of $%
x_{i}$. The network is assumed to form a chain with the nearest neighboring
regulation, $R_{i}\left( x\right) =R_{i}\left( x_{i-1},x_{i+1}\right) $;
however, the formulation presented in this work can be extended to more
complicated cases straightforwardly. The fluctuation may be incorporated
into Eq. (\ref{eq01}) by means of the master equation approach. For this, we
introduce the volume factor $\Omega $ to give the molecular number $n^{\tau
}=\left( n_{1}\text{, }n_{2}\text{, }...\text{, }n_{d}\right) $ as $n^{\tau
}=\Omega x^{\tau }$. In terms of molecular numbers $n$, the corresponding
master equation of Eq. (\ref{eq01}) can be written as 
\begin{equation}
\frac{\partial P\left( n,t\right) }{\partial t}=\sum_{i=1}^{d}\left(
E_{i+}-1\right) \left[ \left( \phi _{i}n_{i}\right) P\left( n,t\right) %
\right] +\Omega \sum_{i=1}^{d}R_{i}\left( x\right) \left[ E_{i-}-1\right]
P\left( n,t\right) ,  \label{eq03}
\end{equation}%
where $P\left( n,t\right) $ is the probability distribution, and the step
operators $E_{i\pm }$ are defined as 
\begin{equation}
E_{i\pm }G\left( n_{i}\right) =G\left( n_{i}\pm 1\right)  \label{eq04}
\end{equation}%
for a function of molecular numbers $G\left( n\right) $. Then, the technique
of $\Omega $-expansion\cite{kampen} is employed to transfer the discrete
process of Eq. (\ref{eq03}) to a continuous process described by the
Fokker-Planck equation, 
\begin{equation}
\frac{\partial \rho \left( x,t\right) }{\partial t}+\nabla \cdot J\left(
x,t\right) =0,  \label{eq05}
\end{equation}%
where $\left( \nabla \right) ^{\tau }=\left( \partial /\partial x_{1}\text{, 
}\partial /\partial x_{2}\text{, }...\text{, }\partial /\partial
x_{d}\right) $, $\rho \left( x,t\right) $ is the distribution density, $%
J\left( x,t\right) $ is the density current defined as \ 
\begin{equation}
J\left( x,t\right) =f\left( x\right) \rho \left( x,t\right) -\frac{1}{\Omega 
}\left[ D\left( x\right) \cdot \partial \right] \rho \left( x,t\right) ,
\label{eq06}
\end{equation}%
and the elements of the diffusion matrix $D\left( x\right) $ are 
\begin{equation}
D_{ij}\left( x\right) =\delta _{i,j}\left[ \frac{R_{i}\left( x\right) +\phi
_{i}x_{i}}{2}\right]  \label{eq07}
\end{equation}%
with the Kronecker delta $\delta _{i,j}=1$ for $i=j$ otherwise $0$, note we
do not sum over repeated indices.

We are interested in the behavior of $\rho \left( x,t\right) $ for the
region near a equilibrium stable point of Eq. (\ref{eq01}), say $x^{\ast }$.
After expanding the density current $J\left( x,t\right) $ of Eq. (\ref{eq06}%
) around the stable point, we obtain the linearized Fokker-Planck equation
for the new variable $y=x-x^{\ast }$ as 
\begin{equation}
\frac{\partial \rho _{L}\left( y,t\right) }{\partial t}+\nabla \cdot
J_{L}\left( y,t\right) =0,  \label{eq008}
\end{equation}%
where $J_{L}\left( y,t\right) $, which contains only the leading order $%
\left( 1/\Omega \right) $ of $J\left( x,t\right) $, is 
\begin{equation}
J_{L}\left( y,t\right) =\left[ F\left( x^{\ast }\right) \cdot y-\frac{1}{%
\Omega }D\left( x^{\ast }\right) \cdot \nabla \right] \rho _{L}\left(
y,t\right) ,  \label{eq009}
\end{equation}%
with the force matrix $F\left( x^{\ast }\right) $ defined as $F_{ij}\left(
x^{\ast }\right) =\left. \partial f_{i}\left( x\right) /\partial
x_{j}\right\vert _{x=x^{\ast }}$, and noting the fact that y itself is of
order $\frac{1}{\omega}$. This leads to a Ornstein-Uhlenbeck process in
which, the drift force is linear and the diffusion is given by a constant
matrix\cite{kampen,uhlenbeck}. The stationary solution of Eq. (\ref{eq008}),
characterized by the condition $\nabla \cdot J_{L}^{S}\left( y\right) =0$,
can be expressed as 
\begin{equation}
\rho _{L}^{S}\left( y\right) =\frac{1}{Z}\exp \left[ -\Phi \left( y\right) %
\right]  \label{eq10}
\end{equation}%
with 
\begin{equation}
Z=\int_{-\infty }^{\infty }\cdot \cdot \cdot \int_{-\infty }^{\infty }\left(
\prod\limits_{m=1}^{d}dy_{m}\right) \exp \left[ -\Phi \left( y\right) \right]
\label{eq10a}
\end{equation}%
and 
\begin{equation}
\Phi \left( y\right) =\frac{1}{2}y^{\tau }\cdot U\left( x^{\ast }\right)
\cdot y,  \label{eq11}
\end{equation}%
where $Z$ can be referred as the partition function, and $U$ is a real
symmetric $d\times d$ matrix\cite{uhlenbeck}. Note that the temperature in
this work is always set to be $1$, $k_{B}T=1$, and hereafter we drop the
arguments for all matrix elements known to be functions of the equilibrium
stable point $x^{\ast }$. One can determine the matrix $U$ by substituting
Eq. (\ref{eq10}) directly into the condition $\nabla \cdot J_{L}^{S}\left(
y\right) =0$ to obtain 
\begin{equation}
tr(F+\frac{1}{\Omega }D\cdot U)-y^{\tau }\cdot (U\cdot F+\frac{1}{\Omega }%
U\cdot D\cdot U)\cdot y=0.  \label{eq12}
\end{equation}%
To solve this for the matrix $U$, we follow an elegant method proposed by Ao%
\cite{ao} and Kwon et al.\cite{kwon} to factorize the force matrix as 
\begin{equation}
F=-\frac{1}{\Omega }\left[ D+Q\right] \cdot U,  \label{eq13}
\end{equation}%
where $D$ is the symmetric diffusion matrix given by Eq. (\ref{eq07}), and $%
Q $ is an antisymmetric matrix which has to be determined. Such
factorization amounts to decomposing the density current into two parts, $%
J_{L}^{S}\left( y\right) =j_{d}^{S}\left( y\right) +j_{c}^{S}\left( y\right) 
$. The first term of Eq. (\ref{eq13}) corresponds to the dissipative part
which generates a motion towards the origin with vanishing density current, $%
j_{d}^{S}\left( y\right) =0$; meanwhile, the second term is the cyclic part
with a divergence-free current density, $j_{c}^{S}\left( y\right) =-\left(
1/\Omega \right) Q\cdot U\cdot y\rho _{L}^{S}\left( y\right) $, which
generates a circulating motion around the constant surface of $\Phi \left(
y\right) $. By substituting Eq. (\ref{eq13}) into Eq. (\ref{eq12}), we
obtain the relation 
\begin{equation}
F\cdot Q+Q\cdot F^{\tau }=F\cdot D-D\cdot F^{\tau },  \label{eq14}
\end{equation}%
which gives enough conditions to determine the matrix $Q$ completely. Thus,
the function $\Phi \left( y\right) $ of Eq. (\ref{eq11}), which can be
referred as the potential energy for the system near a stable point $x^{\ast
}$, becomes 
\begin{equation}
\Phi \left( y\right) =-\frac{\Omega }{2}y^{\tau }\cdot \left[ \left(
D+Q\right) ^{-1}\cdot F\right] \cdot y.  \label{eq15}
\end{equation}

A more intuitive physical picture about the characteristics of the system
may be revealed by mapping $\Phi \left( y\right) $ to the potential energy
of the system of coupled oscillators, 
\begin{equation}
\Phi \left( y\right) =\frac{\Omega }{2}\left[ \sum_{i=1}^{d}\kappa
_{i}y_{i}^{2}+\sum_{i>j}\kappa _{ij}^{c}\left( y_{i}-y_{j}\right) ^{2}\right]
,  \label{eq15-1}
\end{equation}%
which can be casted in the form of 
\begin{equation}
\Phi \left( y\right) =\frac{\Omega }{2}y^{\tau }\cdot V\left( \kappa ,\kappa
^{c}\right) \cdot y.  \label{eq16}
\end{equation}%
Note that though the regulations of the network only come from the nearest
neighbors, the couplings of oscillators may not be restricted to the nearest
neighbors. The force constants, $\kappa $ and $\kappa ^{c}$, can be
specified by equating Eq. (\ref{eq16}) to Eq. (\ref{eq15}), and the
characteristics of the network near a stable point can be expressed in terms
of the force constants. Firstly, based on the partition function of Eq. (\ref%
{eq10a}), which is reduced to 
\begin{equation}
Z=\left( \frac{2\pi }{\Omega }\right) ^{d/2}\left[ \det V\left( \kappa
,\kappa ^{c}\right) \right] ^{-1/2}  \label{eq17}
\end{equation}%
with $\det V\left( \kappa ,\kappa ^{c}\right) $ for the determinant of the
matrix $V\left( \kappa ,\kappa ^{c}\right) $ of Eq. (\ref{eq16}), we may
introduce the effective free energy difference, $\Delta G=-\ln Z$, to
describe qualitatively the steepness of the potential. A stable point with
larger $\Delta G$ value is more easy to focus with less fluctuations.
Furthermore, the variances and covariance of $x_{1}$ and $x_{2}$, defined as 
$\sigma _{i,j}^{2}=\left\langle x_{i}x_{j}\right\rangle -x_{i}^{\ast
}x_{j}^{\ast }$, can be evaluated by using the distribution $\rho
_{L}^{S}\left( y\right) $, 
\begin{equation}
\sigma _{i,j}^{2}=\frac{1}{Z}\int_{-\infty }^{\infty }\cdot \cdot \cdot
\int_{-\infty }^{\infty }\left( \prod\limits_{m=1}^{d}dy_{m}\right)
y_{i}y_{j}\exp \left[ -\frac{\Omega }{2}y^{\tau }\cdot V\left( \kappa
,\kappa ^{c}\right) \cdot y\right] .  \label{eq18}
\end{equation}%
The formulation is applied to two-dimensional regulatory networks, and the
results are given explicitly in the followings.

Consider the case of $d=2$ with regulation functions $R_{1}\left(
x_{2}\right) $ and $R_{2}\left( x_{1}\right) $. The force matrix is 
\begin{equation}
F=\left( 
\begin{array}{cc}
-\phi _{1} & r_{1} \\ 
r_{2} & -\phi _{2}%
\end{array}%
\right) ,  \label{eq20}
\end{equation}%
where $r_{1}$ and $r_{2}$ are defined as $r_{1}=\left. \partial R_{1}\left(
x_{2}\right) /\partial x_{2}\right\vert _{x_{2}=x_{2}^{\ast }}$ and $%
r_{2}=\left. \partial R_{2}\left( x_{1}\right) /\partial x_{1}\right\vert
_{x_{1}=x_{1}^{\ast }}$. Then, the antisymmetric matrix $Q$, determined by
Eq. (\ref{eq14}), is 
\begin{equation}
Q=\left( 
\begin{array}{cc}
0 & W \\ 
-W & 0%
\end{array}%
\right)  \label{eq21}
\end{equation}%
with $W=\left[ r_{2}D_{11}-r_{1}D_{22}\right] /\left( \phi _{1}+\phi
_{2}\right) $. The $F$ and $Q$ matrices given above with the $D$ matrix of
Eq. (\ref{eq07}) yield the potential energy of Eq. (\ref{eq15-1}) as 
\begin{equation}
\Phi \left( y\right) =\frac{\Omega }{2}\left[ \kappa _{1}y_{1}^{2}+\kappa
_{2}y_{2}^{2}+\kappa _{1,2}^{c}\left( y_{1}-y_{2}\right) ^{2}\right] ,
\label{eq22}
\end{equation}%
where the force constants are $\kappa _{1}=\left[ -r_{2}D_{11}+\left( 2\phi
_{1}-r_{1}\right) D_{22}+\left( 2r_{2}-\phi _{2}+\phi _{1}\right) W\right]
/2\left( D_{11}D_{22}+W^{2}\right) $, $\kappa _{2}=\left[ \left( 2\phi
_{2}-r_{2}\right) D_{11}-r_{1}D_{22}-\left( 2r_{1}+\phi _{2}-\phi
_{1}\right) W\right] /2\left( D_{11}D_{22}+W^{2}\right) $, and $\kappa
_{1,2}^{c}=\left[ r_{1}D_{22}+r_{2}D_{11}+\left( \phi _{2}-\phi _{1}\right) W%
\right] /2\left( D_{11}D_{22}+W^{2}\right) $. For the effective free energy
difference, we rescale $\Delta G$ by adding a volume factor, $\Delta 
\overline{G}=$ $-\ln Z-\ln \left[ \Omega /\left( 2\pi \right) \right] $;
then, $\Delta \overline{G}$ becomes half the logarithm of $\det V\left(
\kappa ,\kappa ^{c}\right) $, and it is $\Delta \overline{G}=\left( \frac{1}{%
2}\right) \ln \left[ \kappa _{1}\kappa _{2}+\left( \kappa _{1}+\kappa
_{2}\right) \kappa _{1,2}^{c}\right] $. Moreover, for the potential energy
of Eq. (\ref{eq22}) the variances and covariance of Eq. (\ref{eq18}) become $%
\sigma _{1,1}^{2}=\left[ \left( \kappa _{2}+\kappa _{1,2}^{c}\right) /\Omega %
\right] \exp \left( -2\Delta \overline{G}\right) $, $\sigma _{2,2}^{2}=\left[
\left( \kappa _{1}+\kappa _{1,2}^{c}\right) /\Omega \right] \exp \left(
-2\Delta \overline{G}\right) $, and $\sigma _{1,2}^{2}=\left( \kappa
_{1,2}^{c}/\Omega \right) \exp \left( -2\Delta \overline{G}\right) $. Thus,
expressed in terms of mechanical quantities we summarize the features of the
system implied by the potential energy as follows. In general, the $\Delta 
\overline{G}$ value characterizes the global steepness of the quadratic
potential; the increase of the $\Delta \overline{G}$ value makes the
potential more sharper and, hence, reduces the fluctuations. However, for
the same $\Delta \overline{G}$ value the details of potential shape may have
an effect on the variances and covariance of components; the variance of one
component is proportional to the force constant of the other and to the
coupling strength between the two, and the covariance between the two is
proportional to the coupling strength.

We apply the above results to study the regulation of an auto-regulatory
network of a single gene, which describes the central dogma of gene
expression, transcription and translation. The two variables, $x_{1}$ and $%
x_{2}$, refer to the concentrations of mRNA and protein, respectively. In
this study, we use the most common noise-attenuating regulatory mechanism,
called negative feedback and described by Hill function $R_{1}\left(
x_{2}\right) =k_{\max }/\left[ 1+\left( x_{2}/k_{d}\right) ^{\beta }\right] $%
. Here, $k_{\max }$ is the maximum transcription rate of mRNA, $k_{d}$ is
the binding constant specifying the threshold protein concentration at which
the transcription rate is half its maximum value, and $\beta $ is the Hill
coefficient. \ On the other hand, we set $R_{2}\left( x_{1}\right)
=k_{2}x_{1},$where $k_{2}$ is the translation rate of protein. Then, a
stable equilibrium, $x^{\ast }$, is characterized by two conditions: $\phi
_{1}\phi _{2}-r_{1}\left( x_{2}^{\ast }\right) k_{2}>0$ and $\phi _{1}+\phi
_{2}>0$. Subsequently, one can use Bendixson's criterion to further conlude
that there is no any cycles, only one equilibriuum exists\cite{bendixson}.
For the values of the parameters, we mainly follow those given in Refs. \cite%
{thattai,tao-c}. The half-lifes of mRNA molecules and proteins are set as $2$
minutes and $1$ hour, respectively; this leads to $\phi _{1}=\left( \ln
2\right) /2$ and $\phi _{2}=\left( \ln 2\right) /60$ in the unit of $\left(
\min \right) ^{-1}$.\ The average size of a burst of proteins, $b=k_{2}/\phi
_{1}$, is set as $10$, this leads to $k_{2}=5\left( \ln 2\right) $. By using
the fact that the protein concentration is about $1200$ when $\beta =0$ (no
feedback), we set $k_{\max }=3$\cite{remark}. To study the effect of the
strength of negative feedback on the characteristics of the system, we vary
the parameters $\beta $ from $2$ to $11$, while the $k_{d}$ value is fixed
as $800$.

The numerical results are shown in Fig. 1(a) for the $\kappa _{1}$ and $%
\kappa _{2}$ values and in Fig. 1(b) for the $\kappa _{1,2}^{c}$ and $\Delta 
\overline{G}$ values as functions of the equilibrium concentration of
protein $x_{2}^{\ast }$ for different $\beta $ values. As a consequence of
increasing the $\beta $ value, the $\kappa _{1}$ and $\kappa _{2}$ values
also increase but the $\kappa _{1,2}^{c}$ value decreases; the values are
ranged between $0.305\leq \kappa _{1}\leq 0.354$, $-6.00\times 10^{-4}\leq
\kappa _{2}\leq 6.26\times 10^{-4}$, and $2.74\times 10^{-4}\leq \kappa
_{1,2}^{c}\leq 8.08\times 10^{-4}$, respectively. Note that the $\kappa _{2}$
values are drastically smaller than the $\kappa _{1}$ values, reflecting the
longer half-life of protein. Moreover, the $\Delta \overline{G}$ values for
different $\beta $ given in Fig. 1(b) indicate that the system becomes more
in focus when the $\beta $ value increases.

The fluctuations of the system near a stable point are analyzed by measuring
the variance of $x_{i}$ in terms of the Fano factors $\nu _{i}$, defined as $%
\nu _{i}=\Omega \left( \sigma _{i,i}^{2}/x_{i}^{\ast }\right) $, and the
covariance of $x_{1}$ and $x_{2}$ in terms of the correlation coefficient $%
R_{12}$, defined as $R_{12}=\sigma _{1,2}^{2}/\sqrt{\sigma _{1,1}^{2}\sigma
_{2,2}^{2}}$. The numerical results of $\nu _{1}$, $\nu _{2}$, and $R_{12}$
for different $\beta $ are shown in Fig. 2. The plots indicate that a larger 
$\kappa _{2}$ implies a smaller $\nu _{2}$ and a larger $\kappa _{1,2}^{c}$
implies a larger $R_{12}$. Furthermore, as indicated in the inset of Fig. 2,
the $\nu _{1}$ values all are very closed to but less than one over the
range of $2\leq \beta \leq 11$; the value firstly decreases from $\nu
_{1}=0.9844$ at $\beta =2$, reaches the minimum $\nu _{1}=0.9827$ at $\beta
=4$, and then increases to $\nu _{1}=0.9926$ at $\beta =11$. Because that
the fluctuation of mRNA is caused by a process very close to Poissonian with 
$\nu _{1}=1$, we have very small correlation coefficient ranged between $%
0.015\leq R_{12}\leq 0.101$. We further compare the results with those
obtained from the linear noise Fokker-Planck equation which, as shown
explicitly in Ref.\cite{tao-c}, describes the distribution of fluctuations $%
\xi _{i}\left( t\right) $ introduced via the setting, $x_{i}\left( t\right) =%
\bar{x}_{i}\left( t\right) +\Omega ^{-1/2}\xi _{i}\left( t\right) $, where
the macroscopic values $\bar{x}_{i}\left( t\right) $ are determined by the
rate equations of Eq. (\ref{eq01}). The results thus obtained agree
completely with those shown in Fig. 2. \ 

In conclusion, we present a mechanical viewpoint on the characteristics of
regulatory genetic networks, which is obtained from the energy landscape of
the network near a stable point. The new approach is shown to be consistent
with other descriptions, as demonstrated in the explicit calculations of
auto-regulatory networks, it also provides additional informations, such as
the steepness of a stable point and its relation to fluctuations. Though the
method can also be applied to other genetic networks straightforwardly, it
is limited in the sense that the overall potential landscape cannot be
approximated by a local quadratic approach. For bi-stable systems such as
toggle switches, one might need to patch the potential derived from the two
local minimums in a rigorous fashion to obtain a better result.

\textbf{Acknowledgement}: This work was partially supported by the National
Science Council of Republic of China (Taiwan) under the Grant No. NSC
95-2212-M-033-005 (M.-C. Huang) and the National Science Foundation under
the Grant No.\ PHY-0354776 (Y.-t. Huang).

\newpage 

\FRAME{ftbpFU}{4.5074in}{1.9026in}{0pt}{\Qcb{The force constants, $\protect%
\kappa _{1}$, $\protect\kappa _{2}$, and $\protect\kappa _{1,2}^{c}$, and
the effective free energy difference $\Delta \overline{G}$ for different $%
\protect\beta $ values: $\left( a\right) $ $\protect\kappa _{1}$ (solid
squares) and $\protect\kappa _{2}$ (hollow squares), $\left( b\right) $ $%
\protect\kappa _{1,2}^{c}$ (solid squares) and $\Delta \overline{G}$ (hollow
squares). The horizontal axis is the equilibrium concentration of protein $%
x_{2}^{\ast }$.\ }}{}{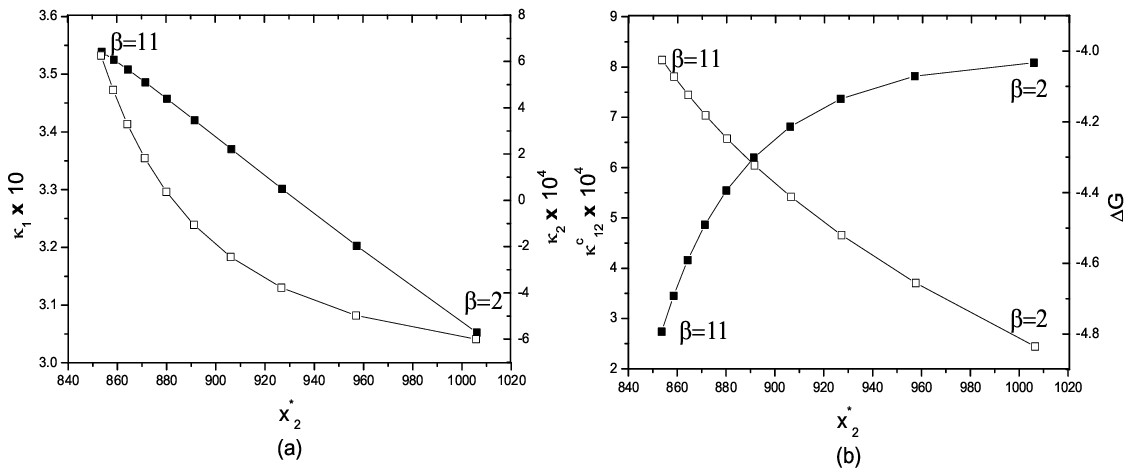}{\special{language "Scientific Word";type
"GRAPHIC";maintain-aspect-ratio TRUE;display "USEDEF";valid_file "F";width
4.5074in;height 1.9026in;depth 0pt;original-width 4.4555in;original-height
1.8645in;cropleft "0";croptop "1";cropright "1";cropbottom "0";filename
'fig1.eps';file-properties "XNPEU";}}\FRAME{ftbpFU}{4.1303in}{3.4696in}{0pt}{%
\Qcb{The Fano factors, $\protect\nu _{1}$ and $\protect\nu _{2}$, and the
correlation coefficient $R_{12}$ for different $\protect\beta $ values. The
horizontal axis is the equilibrium concentration of protein $x_{2}^{\ast }$,
and the inset shows the details of the $\protect\nu _{1}$ values.}}{}{%
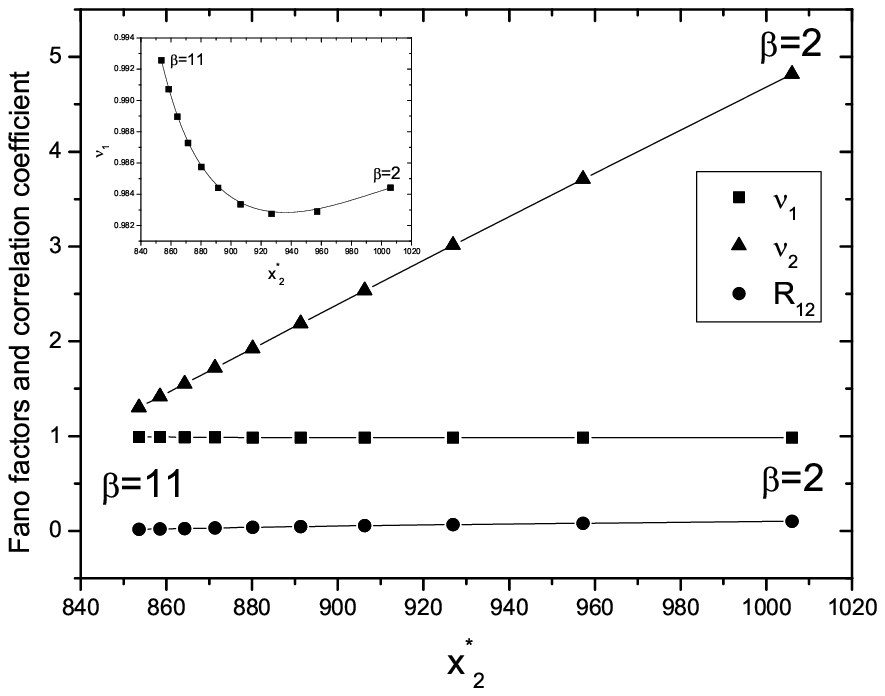}{\special{language "Scientific Word";type
"GRAPHIC";maintain-aspect-ratio TRUE;display "USEDEF";valid_file "F";width
4.1303in;height 3.4696in;depth 0pt;original-width 4.0802in;original-height
3.4238in;cropleft "0";croptop "1";cropright "1";cropbottom "0";filename
'fig2.eps';file-properties "XNPEU";}}


\begin{thebibliography}{99}
\bibitem{remark} The $k_{\max }$ value was set as $2\ln 2\simeq 1.4$ in Ref. 
\cite{thattai}, this corresponds to a mean protein number $600$ when $\beta
=0$ and leads to the increase of mean protein number as the Hill coefficient 
$\beta $ increases. Our setting, $k_{\max }=3$ which is the same as that
given in Ref. \cite{tao-c}, leads to the decrease of mean protein number as $%
\beta $ increases. However, different settings do not change the mechanical
picture obtained in this work.
\end{thebibliography}
\end{document}